\documentclass[apsrev,twocolumn,floats,prd,nofootinbib,superscriptaddress,10pt]{revtex4}

\usepackage{graphicx} 
\usepackage{amsfonts,amsmath,amssymb} 
\usepackage{hyperref}

\usepackage[english]{babel}
\usepackage{verbatim}	
	\usepackage{color}

\begin{document}
	
\title{On the stability of  spherically symmetric  relativistic configurations with scalar field 
}

\author{A.V.~\surname{Tugay}}
\email{tugay.anatoliy@gmail.com}
\affiliation{Taras Shevchenko National University of Kyiv,  Kyiv 01601, Ukraine}
\author{V.I.~\surname{Zhdanov}}
\email{valeryzhdanov@gmail.com}
\affiliation{Taras Shevchenko National University of Kyiv,  Kyiv 01601, Ukraine}
\affiliation{Igor Sikorsky Kyiv Polytechnic Institute, Kyiv,  03056 Ukraine}
\author{Yu.V.~\surname{Taistra}}
\email{ythelloworld@gmail.com}
\affiliation{Pidstryhach Institute for Applied Problems of Mechanics and Mathematics, Lviv, 79060, Ukraine\looseness=-1} 

\begin{abstract}
Static spherically symmetric asymptotically flat configurations in General Relativity with a minimally coupled scalar field typically possess naked singularities. We study a linear stability of these configurations against radial (monopole) perturbations in case of power-law scalar field potentials.  We employ a new shooting method that extracts the eigenvalues corresponding to the exponentially growing  modes of the perturbation. The problem is reduced to one second order ``master'' equation in a divergent form with a regularized effective potential. We confirm our previous results showing that instability modes exist for sufficiently small scalar charges of the initial static configurations.

\end{abstract}


\maketitle{Keywords: General Relativity, scalar fields, naked singularities, linear stability.}

\section{Introduction} \label{Introduction}
\noindent 

High-energy astrophysical phenomena are traditionally associated with black holes. Alternative models may  involve exotic horizonless systems, such as static spherically symmetric (SSS) naked singularities supported by scalar fields (see, e.g., \cite{Joshi_2013, Chowdhury2020} and references therein). It is natural to ask about stability of these objects along with other ``black hole mimickers'' \cite{BH_mimickers2025arXiv250509014B}. 
 
In this paper we focus on the linearized stability SSS configurations  with minimally coupled SFs described by power-law SF potentials. This issue was already addressed in our previous article \cite{TugZhdTai2026}, where the stability problem was reduced to a Schrodinger-like equation with a singular effective  potential. The latter circumstance led to certain difficulties in the numerical integration of the equations. In the present paper we propose a different approach to the numerical investigation of stability based on a different form of the equation for perturbations. 

In Section  \ref{master-intro} we write down the equation for redial perturbations of the SSS configurations and outline some asymptotic properties. All terms in this equation are regular for $x\to 0$. In Section \ref{method-n} we describe the numerical procedure used to find the diverging solutions. In Section \ref{discussion} we summarize the results.

\section{EQUATION FOR PERTURBATIONS}\label{master-intro}
We use the metric of the spherically symmetric space-time in the form
\begin{equation}
    ds^2=  e^\nu dt^2-e^\lambda dr^2-r^2dO^2 
    \label{SS-metric}
\end{equation}
\noindent where 
$ \nu =\nu (\tau, x),\, \lambda =\lambda (\tau, x)$.   

The Einstein equations and equations for minimally coupled with gravity SF follow from the general relativistic action with the  SF potential $V(\phi)=V_0 w(\phi)$ ($V_0>0$) and $\phi w(\phi)>0$. In the numerical simulations we restrict ourselves to the massless case $w(\phi)=\phi^{2n}$ ($n$ is an integer) with $n > 2$; this latter condition ensures the Coulomb asymptotic behavior of the static field at infinity (see, e.g., \cite{ZhdSt}).
Below use the rescaled variables $\tau=\sqrt{V_0}t$, $x=\sqrt{V_0}r$ ($c=G=1$).

In the first stage, we numerically determine the SSS  background solutions $\nu = \nu_0(x)$, $\lambda = \lambda_0(x)$, and $\phi = \phi_0(r)$, which describe asymptotically flat configurations with mass $M$ and scalar field asymptotics $\phi(r) = Q/r$ as $r \to \infty$, where the constant $Q \neq 0$ dubbed  "scalar charge". To do this, we employ the analogous analytical-qualitative and numerical approaches as in \cite{StZhAl2021, SSZ2023quasinormal, TugZhdTai2026}. 

Denote
\begin{equation}
    Y_0=x\left[ {e^{(\nu_0-\lambda_0)/2}}-1\right]
\end{equation}
  and 
\begin{equation}
    Z_0=-x^2 e^{(\nu_0-\lambda_0)/2}\frac{d \phi_0}{d x}=-x(Y_0+x)\frac{d\phi_0}{d x} \,.
\end{equation}
Under above conditions, $|\phi(x)|$ is a monotonically decreasing function on $(0,\infty)$; $Y(x)$ and $Z(x)$ have finite nonzero limits for $x\to 0$, whereas $\phi_0(x)$, $\nu_0(x)$ and $\lambda_0(x)$ have a logarithmic behavior:
\[
\phi_0(x)\sim -\zeta \ln{x}\,,\quad \zeta=\lim\limits_{x\to 0} \frac{Z_0(x)}{Y_0(x)}>0\,, 
\]
\[
\nu_0(x)\sim (\eta-1)\ln {x}\,, \quad \lambda_0(x)\sim (\eta+1) \ln{x}\,, 
\]
where
$
\eta=\frac12 \kappa \zeta^2 $; $\zeta$ is defined numerically depending on the scalar charge $Q$.

In the second stage, after the static solutions are obtained, 
 we look for small radial perturbations $\nu_1(\tau,x)\,, \lambda_1(\tau,x)\,, \phi_1(\tau,x)$ of the static background. 

Linearization of the Einstein-scalar field equations leads to one-dimensional ``master equation''. In the derivation, we omit static terms of the perturbations and express $ \nu_1(\tau,x)$ and  $\lambda_1(\tau,x)$ through $\phi_1(\tau,x)$ to obtain the master equation 
\begin{equation}
A(x)\frac{\partial ^2\phi_1}{\partial \tau^2}=\frac{\partial   }{\partial x}\left(B(x)\frac{\partial \phi_1}{\partial x}\right)-V_{\rm eff}\phi_1 \,,    
\label{master}
\end{equation}
where
\begin{align*}
  A(x)=x^2e^{(\lambda_0-\nu_0)/2}>0\,,  B(x)=x^2 e^{(\nu_0-\lambda_0)/2}>0\,,
\end{align*}
\begin{align}
  V_{\rm eff}=& x^2 e^{\chi_0}\left[ w''(\phi_0)+\kappa x \frac{d\phi_0}{dx}w'(\phi_0)\right]\nonumber \\ & +\kappa x^2 e^{(\nu_0-\lambda_0)/2}\frac{d\phi_0}{dx} \frac{d }{dx}\left(x\frac{d\phi_0}{dx}\right)  \,,\label{Veff}
\end{align}
where $\chi_0=(\nu_0+\lambda_0)/2$; $\kappa=8\pi G/c^4$. 

Using equations for static background,  this expression can be rewritten as
\begin{align}
V_{\rm eff}=& x^2 e^{\chi_0}\left[ w''(\phi_0)-2\kappa  \frac{Z_0}{Y_0+x}w'(\phi_0)\right]\nonumber \\ & +\kappa  e^{\chi_0}\frac{Z_0^2}{(Y_0+x)^2}  \left[ \kappa x^2 w(\phi_0)-1\right]  \,.
\label{Veff++}
\end{align}
The examples of $V_{\rm eff}$ are shown in Figs. \ref{Veff1}, \ref{Veff2}. 

 \begin{figure}[h!]
\centering
\includegraphics[width=.4\textwidth]{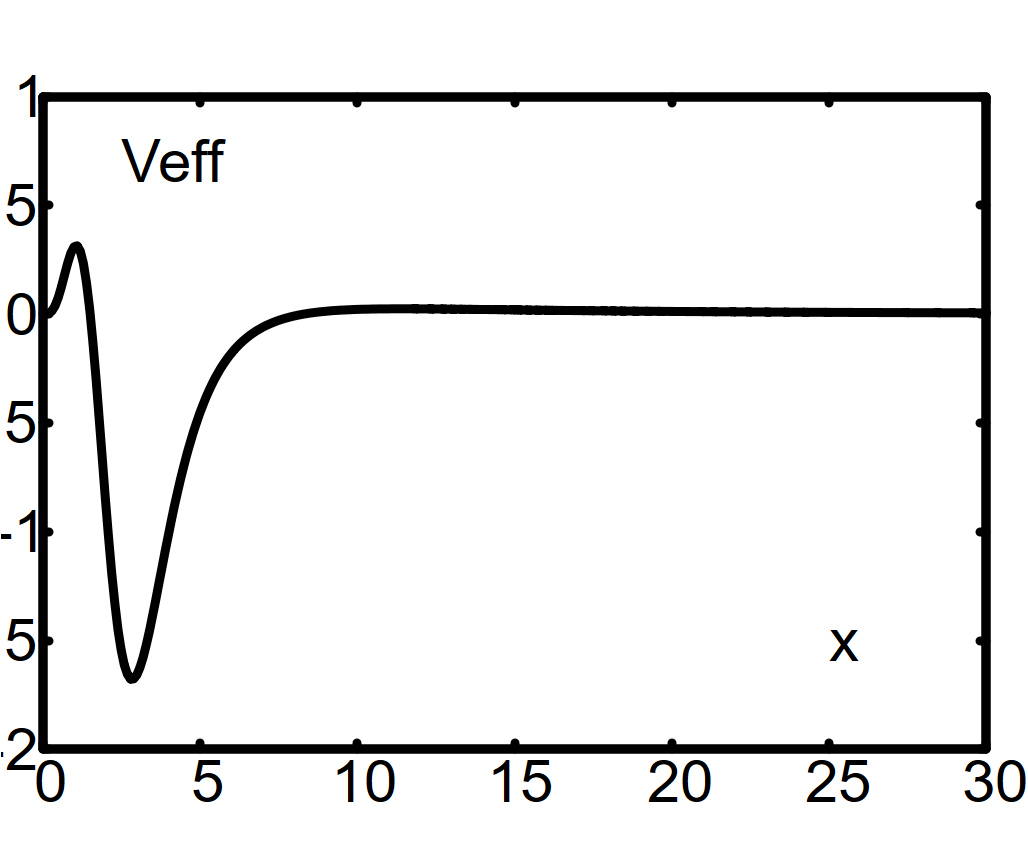}\label{Veff1}
\caption{$V_{eff}(x)$ for $Q=1< Q_{max}$ and $n=3$.} 
\label{Veff1}  
\end{figure}

 \begin{figure}[h!]
\centering
\includegraphics[width=.4\textwidth]{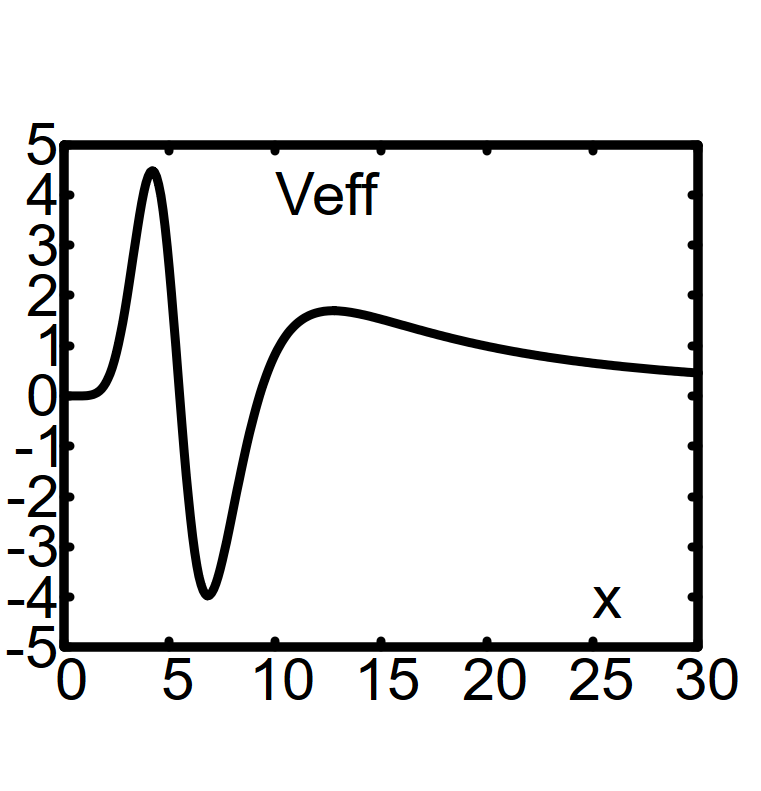}\label{Veff2}
\caption{$V_{eff}(x)$ for $Q=2>Q_{max}$ and n=3.} 
\label{Veff2}  
\end{figure}

For $x\to \infty$ we have $A(x)\approx x^2$, $B(x)\approx x^2$, $Y(x)\to -r_g$ ($r_g=2MG/c$); $Z(x)\to Q$,  $\chi_0(x)\to 0$; then $V_{\rm eff}\sim x^{-2}(30 Q^4-\kappa Q^2)$ for $n=3$ and  $V_{\rm eff}\sim -x^{-2}\kappa Q^2$ for $n>3$.  

Representation (\ref{master}) of the master equation differs from that in \cite{TugZhdTai2026}; here the potential $V_{\rm eff}$ is free from singular terms for $x\to 0$; this is essential for numerical calculations near the center; also $A(x)\sim x^3$ and $B(x)\sim x$ for $x\to 0$. This facilitates the numerical investigation of the problem.

Due to spherical symmetry, monopole perturbations depend only on the radial coordinate $x$ and time $\tau$. We look for  a perturbation in the form:
\begin{equation}
 \phi_1(x,\tau) = \Phi(x) e^{-i\omega \tau},
\end{equation}
where real  $-i\omega =\Omega > 0$ corresponds to exponentially growing (unstable) modes ($\sim e^{\Omega \tau}$). This yields 
\begin{equation}
\frac{d}{dr} \left( B(r) \frac{d\Phi}{dr} \right) - V_{\text{eff}}(r) \Phi(r) = \Omega^2 A(r) \Phi(r),
\label{eq:master-x}
\end{equation}
In case of positive $Q^2$, asymptotics for $x\to\infty$ 
\begin{equation}
  \Phi(\Omega,x)= x^{-1}\left[ J_{-}(\Omega)e^{-\Omega x}+J_{+}(\Omega) e^{\Omega x} \right]
\end{equation}
with constants $J_\pm(Q)$. 

For $x\to 0$ the general solution is 
\begin{equation}
  \Phi(\Omega,x)= C_1+C_2 \ln{x}
\end{equation}
To determine whether unstable modes exist, we formulate an eigenvalue problem with boundary conditions ensuring regular behavior at the center  ($C_2=0$, $\Phi(\Omega,x)\simeq C_1+O(x^{\eta+1})$) and  integrability due to sufficiently rapid decay at spatial infinity ($\Phi\to 0$ as $x \to \infty$). In this space of functions, operator of the left-hand side of (\ref{eq:master-x}) is symmetric yielding the real eigenvalue $\Omega^2$. In this work, the primary focus is on positive values of $\Omega^2$ corresponding to unstable modes.

\section{NUMERICAL  METHOD}\label{method-n}
The numerical procedure consists of solving equation  (\ref{eq:master-x}) with regular conditions for $x\to 0$ ($\Phi(x)\to 1$, $\Phi'(x)\to 0$), which fix the initial Cauchy data for the numerical integration up to an arbitrary normalization constant. Using this solution, we study the function 
 \[ \Phi_1(\Omega,x)\triangleq
 \Phi(\Omega,x)xe^{-\Omega x} \,,   \]
  which for large $x$ can be approximated as 
  \[
  \Phi_1(\Omega,x)=   J_{-}(\Omega)e^{-2\Omega x}+J_{+}(\Omega)
  \]
  and approaches   constant $J_{+}(\Omega)$. By changing $\Omega$ we find the root of equation  
\begin{equation}\label{Omega}
J_{+}(\Omega)=\lim\limits_{x\to\infty}\left[\Phi(\Omega,x)xe^{-\Omega x}\right]=0\,;
\end{equation}
here $\Omega$ is the eigenvalue we are looking for. In this case we have the necessary asymptotic behavior of solution of (\ref{eq:master-x}) at infinity $\Phi(x)\sim  x^{-1} e^{-\Omega x} $, which is at the same time regular at the center. 
 
To find a numerical solution of (\ref{Omega}) we scan an interval of admissible $Q$-values to look for different  $Q_1,Q_2$ corresponding to  opposite signs of $J_+(Q_1)$ and  $J_+(Q_2)$ and then apply the bisection
algorithm yielding the solution. It is important to note that to prove an instability of configuration with some scalar charge $Q$, it is sufficient to find at least one divergent solution with $Q^2>0$. We found such divergent solutions for each $n$, but for sufficiently small $Q<Q_{max}$, where the critical values $Q_{max}$ depend on $n$. The results are in satisfactory agreement with our previous results \cite{TugZhdTai2026}.

\begin{figure}[h!]
\centering
\includegraphics[width=0.45\textwidth]{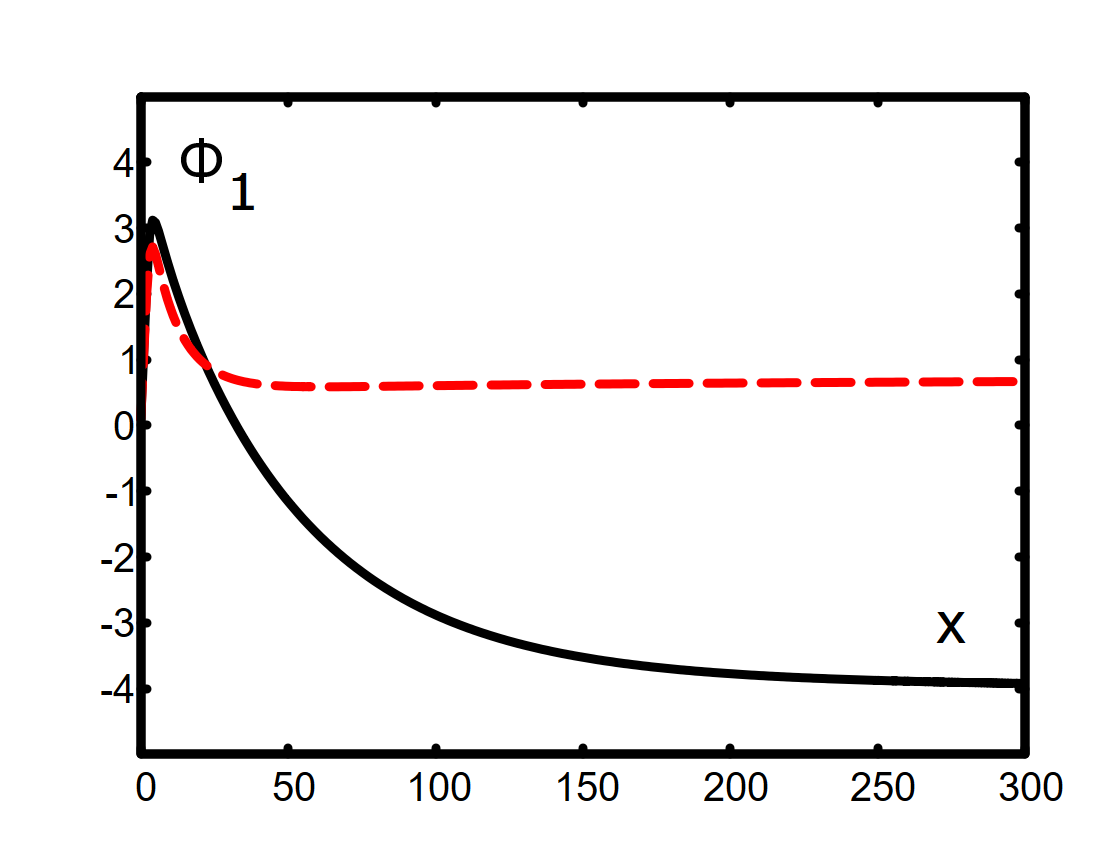}
\caption{The profile of $\Phi_1(x)$ for two trial values of $\Omega ^2$: 0.0001 for black solid line and 0.002 for red dashed line. These profiles constrain the exact eigenvalue $\Omega ^2=0.00096$ for Q=1 and n=3.}
\label{fig:Phi_1}
\end{figure}

Example of $\Phi_1(\Omega,x)  $ for two values of $\Omega$ is shown on  Fig. \ref{fig:Phi_1}. 

\begin{figure}[h!]
\centering
\includegraphics[width=0.45\textwidth]{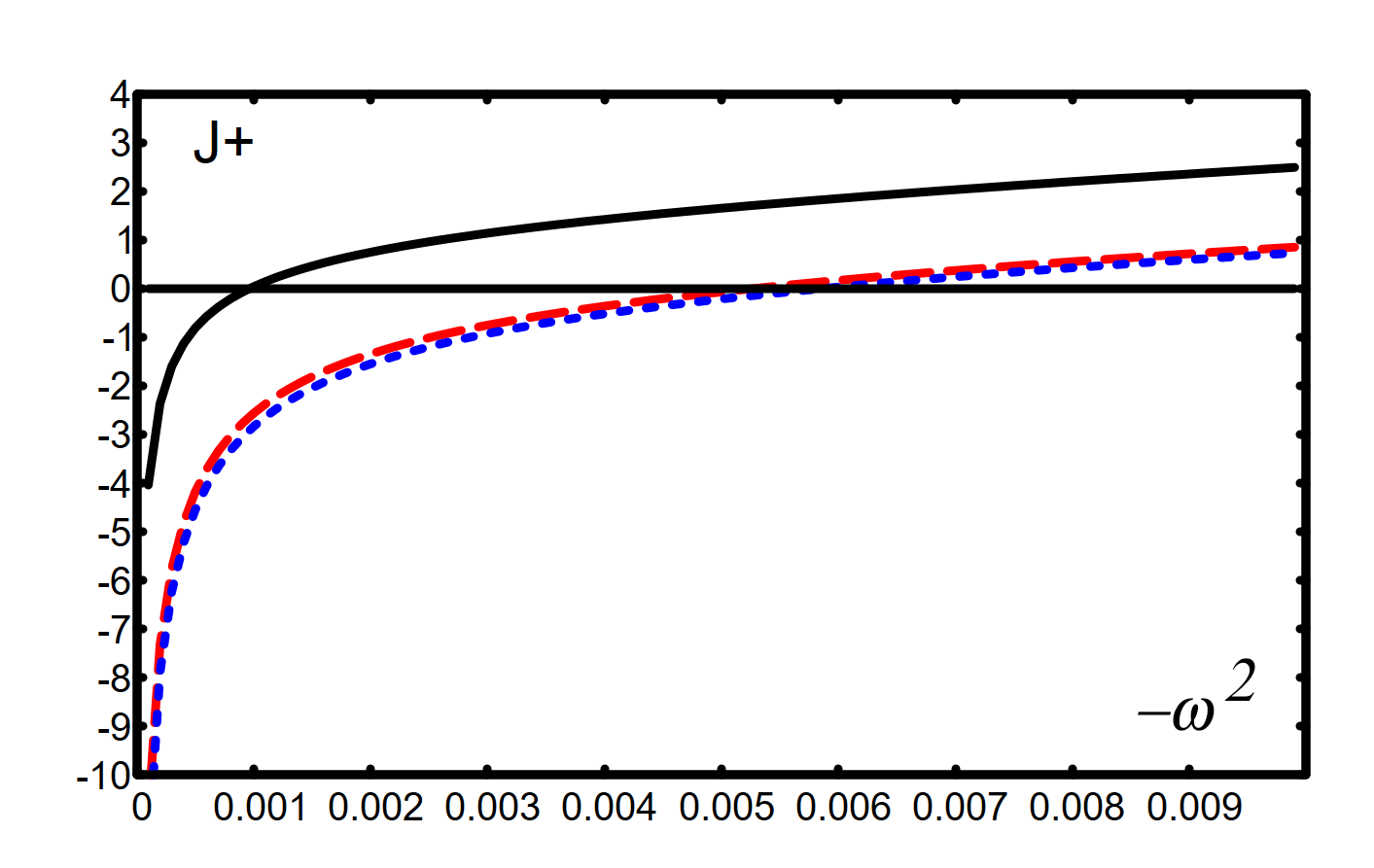}
\caption{Dependence of asymptotic limits $J_+$ of normalized scalar field perturbation $x\Phi(\Omega,x)e^{-\Omega x}$ from  $\Omega^2=-\omega^2$ for SF potentials with $n=3$ (black solid line), $n=4$ (red long dashes) and $n=5$ (blue short dashes); Q=1. The points of intersection of the curves with the x-axis provide the solution to equation (\ref{Omega}). }
\label{fig:Jplus}
\end{figure}

\begin{figure}[h!]
\centering
\includegraphics[width=0.45\textwidth]{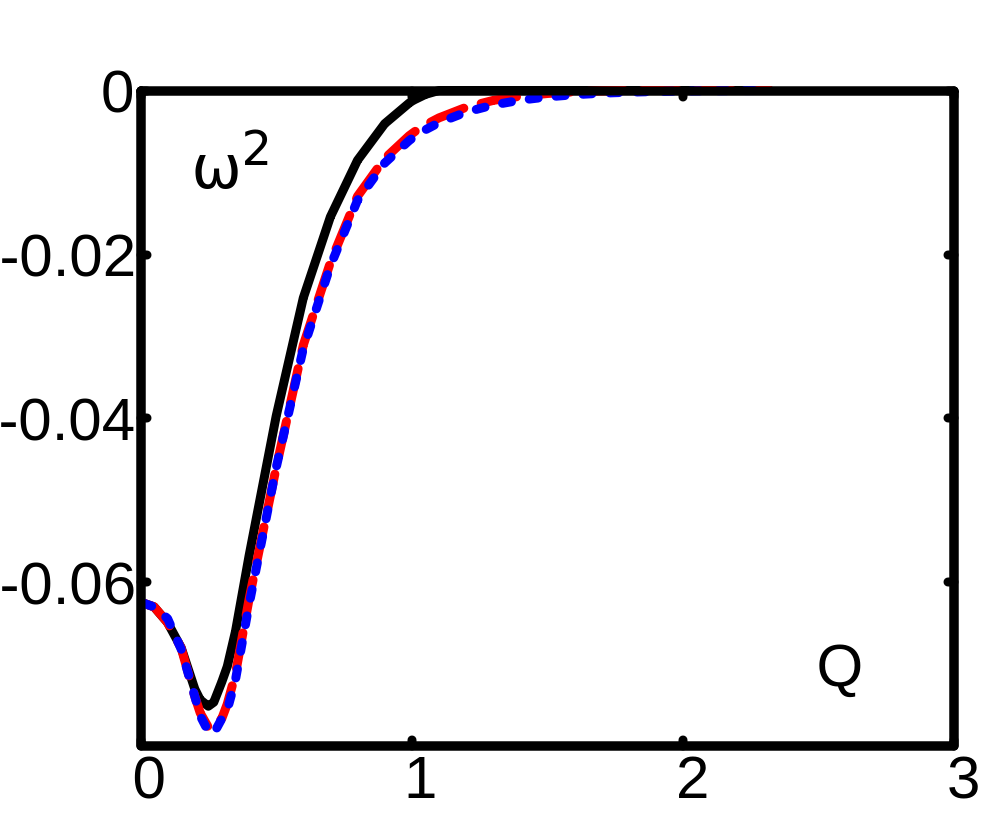}
\caption{Dependence of eigenvalues $\omega^2=-\Omega^2$ on the scalar charge Q for SF potentials with $n=3$ (black solid line), $n=4$ (red dashes) and $n=5$ (blue dashes). }
\label{fig:Y2}
\end{figure}





\section{CONCLUSIONS}\label{discussion}
We investigated static spherically symmetric configurations in General Relativity with one real minimally coupled SF.  We used   massless SF potential  $V_0 \phi^{2n}$ ($n>2$ is an integer,  $V_0>0$) and $\phi w(\phi)>0$. 
The main attention was paid to stability against radial perturbations. To this end, we employed a numerical method different from that used in \cite{TugZhdTai2026}; its advantage lies in the fact that it allows one to avoid singularities of the effective potential in the equation for perturbations. The results obtained are fully consistent with our previous ones \cite{TugZhdTai2026}.

In the present paper, we detailed the implementation and mathematical foundation of the shooting method used to find unstable modes of radial perturbations radial eigenvalues of naked singularities supported by power-law scalar fields.  

Key findings of this analysis include:
\begin{enumerate}
    \item The asymptotic function $J_+(\Omega)$ serves as a reliable numerical indicator for unstable bound states, exhibiting a sign change across true eigenvalues.
    \item Unstable radial modes ($\omega^2 < 0$) are present for small scalar charges $Q < Q_{{max}}$, confirming the physical instability of weak-field naked singularities. This supports the cosmic censorship hypothesis.
    \item However, for large values of the scalar field charge ($Q > Q_{{max}}$), we have not found the  unstable eigenvalues, demonstrating that strong scalar fields may stabilize static spherically symmetric naked singularities against linear monopole perturbations.
\end{enumerate}

\vskip3mm \textit{Acknowledgements.}
This work was supported in part by the Ministry of Education and Science of Ukraine (project 0124U001660).


\end{document}